\documentclass[10pt,journal,compsoc]{IEEEtran}
\usepackage{graphicx}
\usepackage{booktabs} 
\usepackage{array}
\newcolumntype{P}[1]{>{\centering\arraybackslash}p{#1}}
\ifCLASSOPTIONcompsoc
  \usepackage[nocompress]{cite}
\else
  \usepackage{cite}
\fi
\usepackage{multirow}
\ifCLASSINFOpdf
\else
\fi
\usepackage{multirow}
\usepackage{amsmath}
\usepackage{acronym}
\usepackage{algorithmic}
\usepackage{subfigure}

\begin{document}

\title{Developing a Multi-variate Prediction Model For The Detection of COVID-19 From Crowd-sourced Respiratory Voice Data}

\author{Wafaa Aljbawi, Sami O. Simmons, and
        Visara Urovi
}


\markboth{February~2022}
{Shell \MakeLowercase{\textit{et al.}}: Bare Advanced Demo of IEEEtran.cls for IEEE Computer Society Journals}

\IEEEtitleabstractindextext{%
\begin{abstract}
COVID-19 has affected more than 223 countries worldwide. There is a pressing need for non-invasive, low-costs and highly scalable solutions to detect COVID-19, especially in low-resource countries where PCR testing is not ubiquitously available. Our aim is to develop a deep learning model identifying COVID-19 using voice recordings data spontaneously provided by the general population (voice recordings and a short questionnaire) via their personal devices. The novelty of this work is in the development of a deep learning model for the identification of COVID-19 patients from voice recordings.

\textbf{Methods:} 
We used the Cambridge University dataset consisting of 893 audio samples, crowd-sourced from 4352 participants that used a COVID-19 Sounds app. Voice features were extracted using a Mel-spectrogram analysis. Based on the voice data, we developed deep learning classification models to detect positive COVID-19 cases. These models included Long-Short Term Memory (LSTM) and Convolutional Neural Network (CNN). We compared their predictive power to baseline classification models, namely Logistic Regression and Support Vector Machine.

\textbf{Results:} LSTM based on a Mel-frequency cepstral coefficients (MFCC) features achieved the highest accuracy (89\%,) with a sensitivity and specificity of respectively 89\% and 89\%, The results achieved with the proposed model suggest a significant improvement in the prediction accuracy of COVID-19 diagnosis compared to the results obtained in the state of the art.

\textbf{Conclusion:} 
Deep learning can detect subtle changes in the voice of COVID-19 patients with promising results. 
As an addition to the current testing techniques this model may aid health professionals in fast diagnosis and tracing of COVID-19 cases using simple voice analysis.
\end{abstract}

}

\maketitle

\IEEEdisplaynontitleabstractindextext
\IEEEpeerreviewmaketitle

\ifCLASSOPTIONcompsoc
\IEEEraisesectionheading{\section{Introduction}\label{sec:introduction}}
\else
\section{Introduction}
\label{sec:introduction}
\fi
\IEEEPARstart{T}{he} coronavirus or severe acute respiratory syndrome coronavirus 2 (SARS-CoV-2) can cause severe respiratory disease in humans and has become a potential threat to human health and the global economy. On February 1, 2022, more than 5.5 million deaths from coronavirus disease 2019 (COVID-19) were confirmed \cite{worldometer2022covid}. Since the outbreak in late December 2019, it has spread worldwide \cite{lai2020severe}.
One of the most concerning aspects of SARS-CoV-2 is its rapid spread; the virus easily spreads through surfaces \cite{for2021science}, air, breathing, talking or physical contact \cite{ningthoujam2020covid}, and thus it is possible to affect large populations in a very short time. As a result, it is important to quickly identify who is infected. Several research efforts have been carried out in order to avoid the rapid spread of the epidemic and to effectively control the number of infected people \cite{han2022sounds},\cite{stasak2021automatic},  and\cite{de2021covid}. In order to facilitate the detection of cases, several researchers have been exploring the possibility of utilizing auditory data produced by the human body (such as breathing \cite{hassan2020covid}, heart rate \cite{mehrabadi2021detection}, and vibration sounds \cite{liang2017vibration}) to diagnose and track disease progression \cite{brown2020exploring}.

The most common symptoms of the virus include fever, dry cough, loss of smell and taste, headache, muscle aches, diarrhea, conjunctivitis, and in more severe cases, shortness of breath, chest pain, and loss of speech or movement \cite{vahedian2021you}. The respiratory track is impacted which can implicitly impact the voice of the affected patients by resulting in a lack of speech energy and a loss of voice due to shortness of breath and upper airway congestion \cite{despotovic2021detection}. Recurrent dry coughs can also cause alterations in the vocal cords, reducing the quality of one's voice. A recent study found that individuals with COVID-19 had changes in their voice's acoustic characteristics due to inadequate airflow through the vocal tract as a result of pulmonary and laryngological dysfunction \cite{despotovic2021detection}. As a consequence, COVID-19 as a respiratory condition may cause patients' voices to become distinctive, leading to distinguishable voice signatures.

With the emergence of COVID-19, several attempts were made to complement standard testing procedures with effective automated diagnosis solutions. The World Health Organization (WHO) considers nucleic acid-based real time reverse transcription polymerase chain reaction (RT-PCR) to be the standard technique for diagnosing COVID-19 \cite{vahedian2021you}. Due to the high cost and restricted access to the test, as well as the risk of exposing healthcare professionals and medical staffs to the virus, it is also infeasible to go to medical centers and undergo RT-PCR after every cough or uncomfortable feeling \cite{vahedian2021you}. The Rapid Antigen Test (RAT) is an alternative that does not require laboratory processing and eliminates the time constraint of RT-PCR; however, its sensitivity declines with lower viral loads, resulting in false negative results in patients with lower levels of the SARS-CoV-2 virus \cite{despotovic2021detection}. As a result, finding novel accessible and non-invasive techniques for diagnosing this disease with greater precision is critical, as it can alleviate unnecessary concerns and motivates patients to take necessary measures.

In this research, we developed an AI-based algorithm that identifies patients who report being COVID-19 positive
based on the results of their voice analysis. 
Several other works have been proposed to tackle this problem such as \cite{han2022sounds},\cite{stasak2021automatic},\cite{de2021covid},\cite{nassif2022covid},\cite{chang2021covnet}, and \cite{aly2022pay}. The novelty of our work is that we trained our models on a large multi-modal dataset of COVID-19 respiratory sounds. Besides, we developed both deep learning classification models and basic machine learning models to determine which model is optimal for diagnosing COVID-19.

The reminder of the paper is organized as follows. Section \ref{background} provides a literature review on voice analysis for COVID-19 diagnosis. The architecture of the proposed models and the used dataset in this research are explained in detail in section \ref{methods}. Section \ref{exp} describes the features used to train our models and other preprocessing procedures. Section \ref{expeiments} explains the experiments that are conducted to evaluate the performance of the used models. This is followed by a description of the results and their interpretation in section \ref{results}. Finally, section \ref{con} summarises the work and identifies future directions.

\section{Background and Related Work}
\label{background}

Voice samples provide a plethora of health-related information \cite{fagherazzi2021voice}. As a result, scientists believe that minor voice signals might reveal underlying medical issues or disease risk \cite{fagherazzi2021voice}. Voice analysis technologies have the potential to be a reliable, efficient, affordable, convenient, and simple-to-use technique for predicting, diagnosing, and treating health problems. Various approaches are used to extract certain voice and acoustic features (referred to as vocal biomarkers) from audio voice samples. These features are then analyzed for important patterns and cues to give insights about an individual's health.

Mild or severe changes in one's voice can be a sign of a variety of diseases, making vocal biomarkers a noninvasive tool to monitor patients, grade the severity, the stages of a disease or for drug development \cite{fagherazzi2021voice}. Patients suffering from Parkinson's disease, for example, have a decrease of voice volume and start talking more quickly \cite{solana2021analysis}. Patients with Alzheimer's disease experience longer pauses between words, have difficulty finding their words, and frequently substitute nouns with pronouns like "that" and "it" \cite{almor1999alzheimer}. As a result, modifications in voice, which are normally undetectable by human ears, are now being investigated in order to identify early indicators of health problems. For instance, in 2019, researchers successfully implemented a model for diagnosing Parkinson’s disease \cite{wroge2018parkinson}. Their model was able to distinguish between Parkinson's disease patients and those in the control group 85\% of the time \cite{wroge2018parkinson}. Interestingly, the proposed model outperformed the clinical diagnosis set by nonspecialist doctors (with accuracy of 74\% at the time) and diagnosis set by disorder experts (with accuracy on average 80\%) \cite{wroge2018parkinson}.

Similarly, several other studies have proposed automatic COVID-19 screening by analyzing coughs or respiratory signals, to detect the presence of alterations due to the COVID-19 infection. A few other studies, instead, have indicated solutions for the detection of COVID-19 disorders based on an analysis of voice samples. To illustrate, in \cite{schuller2021interspeech}, the authors propose a technique for identifying this illness based on analyzing a specific sentence ("I believe my data may assist to manage the virus pandemic."). This sentence was repeated 1-3 times and was reported along with the COVID-19 test results. Cambridge University shares this dataset, which is crowdsourced from more than 10 different countries including samples from 366 subjects with 308 COVID-19 positive patients, and 893 recordings. The authors achieved an AUC of 72.1\% for distinguishing COVID-19 subjects from non-COVID-19 subjects using Support Vector Machine (SVM) algorithm on COMPARE Acoustic Feature Set that contains 6373 static features resulting from the computation of functionals (statistics) over low-level descriptor (LLD) contours.

Besides, researcers have investigated several methods to extract sound features. Mel Frequency Cepstrum Coefficient (MFCC) is a technique for extracting audio features that is extensively utilized in different audio recognition applications such as speech emotion identification and abnormal voice recognition. In fact the most important point about speech is that the sounds produced by a human are filtered by the shape of the vocal tract, which includes the tongue, teeth, and so on. This shape controls how the sound is produced. The shape of the vocal tract reveals itself in the envelope of the short time power spectrum, and the goal of MFCCs is to appropriately capture this envelope. Moreover, a recent study showed the remarkable proficiency of MFCCs in detecting COVID-19 \cite{nassif2022covid}. The researchers have extracted MFCC features from coughing, breathing, and speaking sounds using a dataset of 1159 sound samples obtained from 592 participants with 213 COVID-19 infected patients \cite{nassif2022covid}. These were then processed using the Long Short-Term Memory (LSTM) architecture, which achieved 98.9\% accuracy, precision, F1-score, and recall \cite{nassif2022covid}.

With the development of deep learning and machine learning, neural networks have played an important role in audio recognition. For instance, LSTM \cite{hochreiter1997long}, SVM \cite{cortes1995support}, CNN \cite{o2015introduction}, ANN \cite{mcculloch1943logical}, KNN \cite{altman1992introduction}, and other methods are widely used for speech recognition. Compared with traditional methods, deep and machine learning models can extract and learn more complex and robust features and make intelligent decisions on its own.

In conclusion, there are relatively few studies in the literature related to the identification of COVID-19 using an analysis of speech sounds, most likely due to the virus's recent spread and the pandemic's ongoing development. The majority of these efforts have been conducted on relatively small and, in many cases, inaccessible datasets, limiting their accessibility to the larger research community and limiting the future development of classification algorithms on standardized datasets.

\section{Methods}
\label{methods}

\begin{figure*}[h!]
   \centering
  \includegraphics[width=\textwidth,height=10cm]{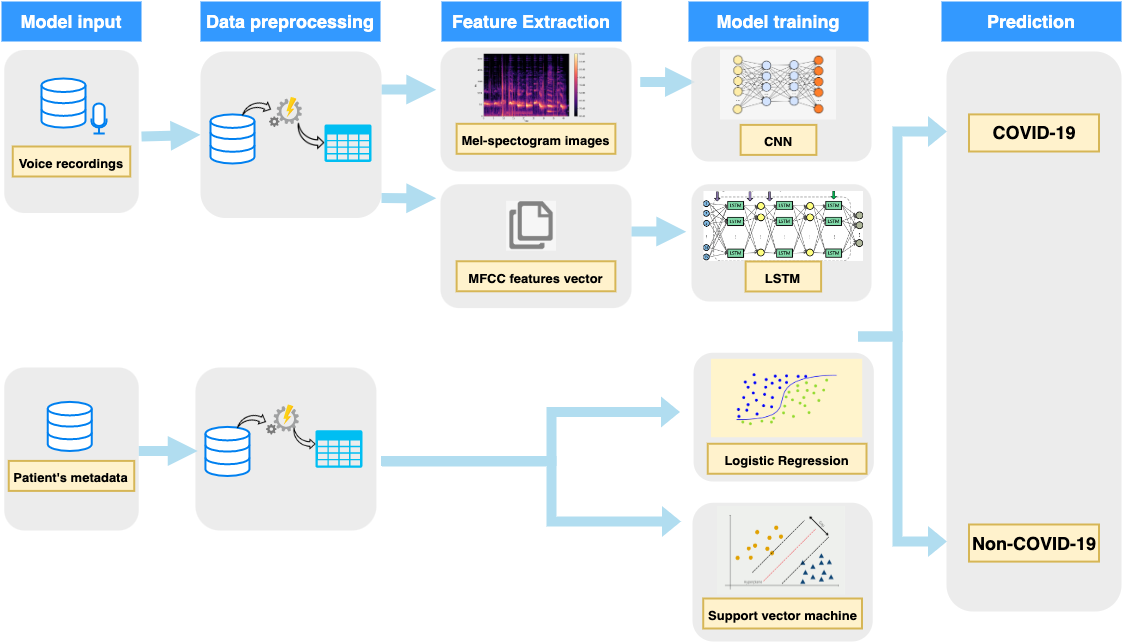}
  \caption{The used pipeline for both basic Machine learning classifiers and Deep Learning classifiers for COVID-19 binary classification (e.g., COVID-19 vs. non-COVID-19). }
  \label{fig:pipeline}
\end{figure*}

In this section we will describe the data used for this analysis, the steps followed to clean and the pre-processing performed prior to the analysis. Figure \ref{fig:pipeline} illustrates the model development steps into a pipeline.

\subsection{Dataset Description}
\begin{figure*}[htp]
  \centering
  \subfigure[Smoking status of the participants]{\includegraphics[scale=0.30]{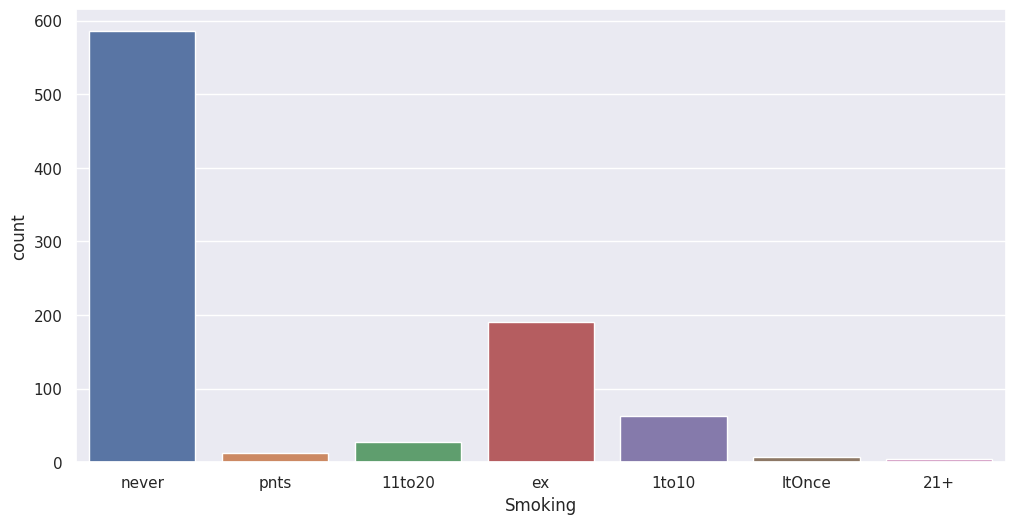}}\quad
  \subfigure[The number of participants who admitted to the hospital]{\includegraphics[scale=0.30]{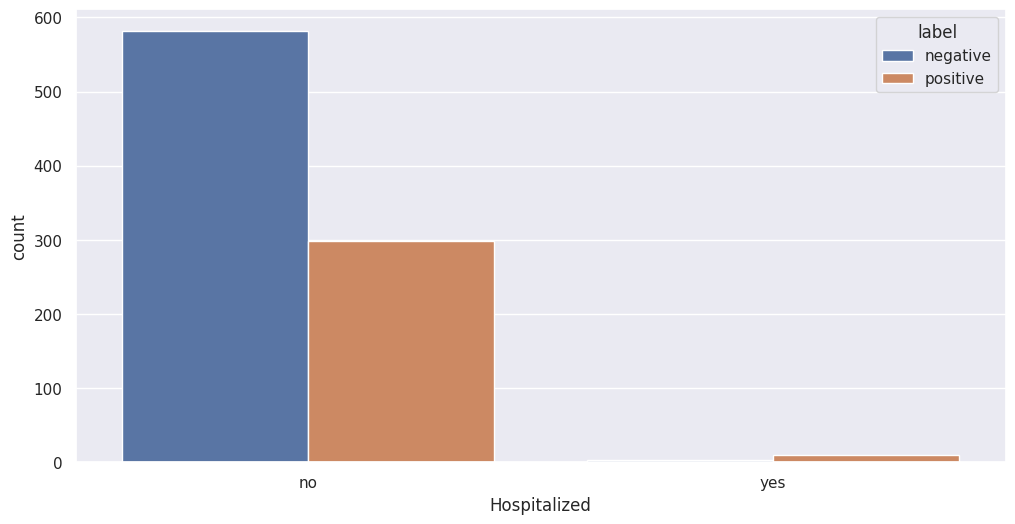}}\quad
  \subfigure[A COVID-19 test result]{\includegraphics[scale=0.30]{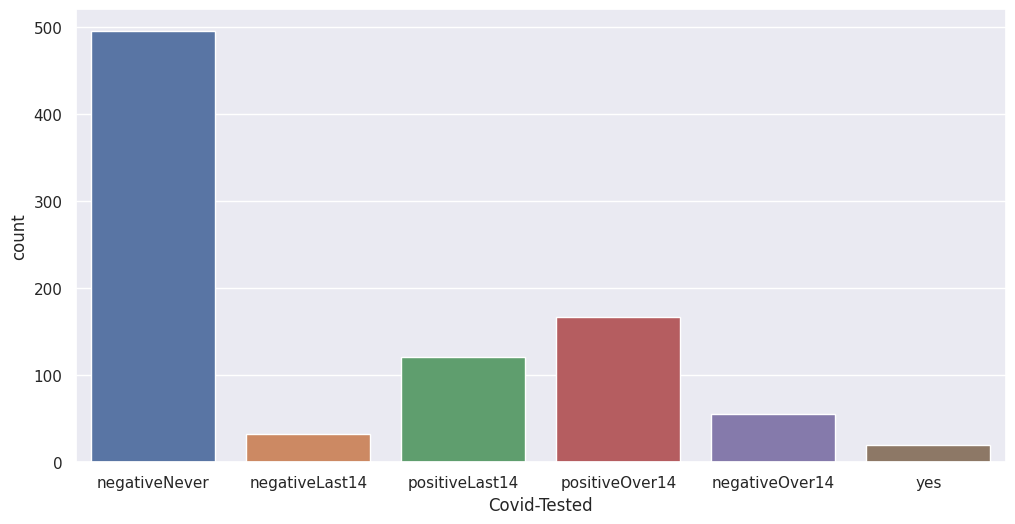}}
  \caption{Users characteristics (a) age, (b) gender, (c) smoking status, (d) hospitalized or not (e) their COVID-19 test result }
  \label{fig:user}
\end{figure*}

In this work we used the University of Cambridge dataset \cite{brown2020exploring} which was crowd-sourced and collected from a web-based platform, an android application, and an iOS application. As reported by \cite{brown2020exploring}, participants are asked to report their demographics, medical history, and smoking status. In addition, they are required to report their COVID-19 test results, hospitalisation status, and symptoms (if any). To record respiratory sounds, participants are prompted to i) cough three times; ii) breathe deeply through their mouth three to five times; iii) read a short sentence on-screen and repeat it three times. After a year of data collection, 4352 users participants worldwide have contributed their data, and 893 audio samples (308 COVID-19 positive samples) are released. In Figure \ref{fig:user}, user characteristics are shown. The average age of our participants is 46 where the max and min age are 90 and 9.5 years old, respectively. The dominant gender is male with 503 males 385 females participants. Also, most of the participants (593 out of 893) are non-smokers. However, the majority of the participants don't have a medical condition but high blood pressure is so common among other participants. Finally the symptoms they suffered from include dry cough, smell and taste loss, sore throat, wet cough, short breath, muscle ache, and headache.

The training data set was divided into training and testing sets. 70\% of cases were selected randomly and were considered as the training set, and the rest were designated as the testing set. Thus, the training set size was 319, and the testing set size was 138.

Our goal is to determine whether a patient has COVID-19 or not based on the voice data. Therefore, we developed several classification models to detect positive COVID-19 cases. These models included Long-Short Term Memory (LSTM) and Convolutional Neural Network (CNN). We compared their predictive power to baseline classification models, namely Logistic Regression and Support Vector Machine.
The following section explains the architecture of the suggested models.

\begin{figure*}[htp]
  \centering
  \subfigure[Age distribution of the participants]{\includegraphics[scale=0.38]{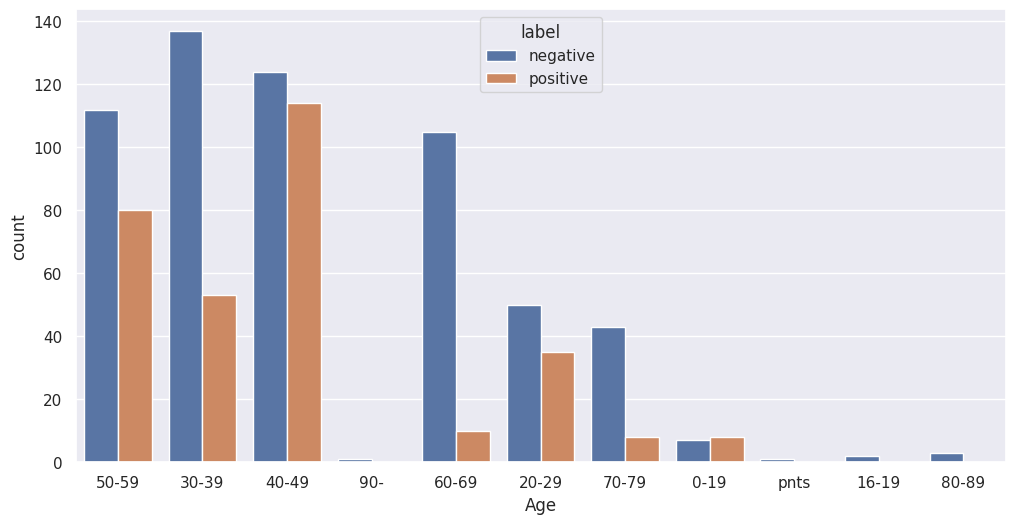}}\quad
  \subfigure[Gender distribution of the participants]{\includegraphics[scale=0.50]{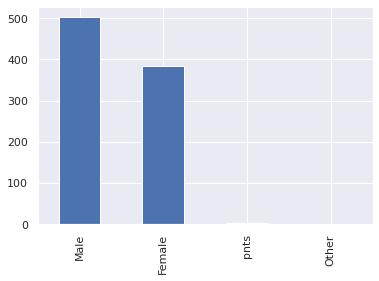}}
\end{figure*}

\begin{table*}[t]
  \centering
\begin{tabular}{|c|c|c|c|c|c|c|}
\hline
Attribute & Variable & Value & Count & Percentage  \\ 
\hline
\multirow{2}{*}{Demographics} & \multirow{3}{*}{Gender}  & Female & 385 & 43\%  \\ 
\cline{3-5}
& & Male & 503& 56\% \\
\cline{3-5}
& & Other & 5 &  0.6\% \\
\hline
 &\multirow{3}{*}{Age} & Max age & 90 &   \\ 
\cline{3-5}
& & Min age & 9.5 &  \\
\cline{3-5}
& & Avg age & 46 &   \\
\hline
\multirow{9}{*}{Symptoms}& & None & 375 & 42\% \\ 
\cline{3-5}
& & Dry cough & 118 & 13\% \\
\cline{3-5}
& & Smell and taste loss, Sore throat & 47 & 5.3\%  \\
\cline{3-5}
& & Wet cough & 37 &  4.1\% \\
\cline{3-5}
& & Short breath & 25 & 2.8\%  \\
\cline{3-5}
& & Sore throat & 22 & 2.5\%  \\
\cline{3-5}
& & Muscleache & 15 & 1.7\%  \\
\cline{3-5}
& & Headache & 14 &  1.6\% \\
\cline{3-5}
& & Dry cough, Tightness  &12 &1.3\%\\
\hline
\multirow{6}{*}{Prior Medical History}& & None & 630 & 71\% \\ 
\cline{3-5}
& & High blood pressure & 112 & 13\% \\
\cline{3-5}
& & Pulmonary & 50 & 6\%  \\
\cline{3-5}
& & Other & 35 &  4\% \\
\cline{3-5}
& & Diabetes & 23 & 2.6\%  \\
\cline{3-5}
& & Cardiovascular & 4 & 0.4\%  \\
\hline
\multirow{7}{*}{Smoking Status}& & Non-smoker & 586 & 71\% \\ 
\cline{3-5}
& & Ex-smoker & 191 & 13\% \\
\cline{3-5}
& & 1-10 cigarettes per day & 63 & 6\%  \\
\cline{3-5}
& & 11-20 cigarettes per day & 28 &  4\% \\
\cline{3-5}
& & Prefer not to say & 13 & 2.6\%  \\
\cline{3-5}
& & Smoked Once & 7 & 0.4\%  \\
\cline{3-5}
& & 21+ cigarettes per day & 5 & 0.4\%  \\
\hline
\multirow{2}{*}{Hospitalized}& & Yes & 880 & 98.5\% \\
\cline{3-5}
& & No & 12 & 1.5\% \\
\hline
\multirow{2}{*}{COVID-19 test result}& & negative & 585 & 65.5\% \\
\cline{3-5}
& & positive & 308 & 34.5\% \\
\hline
\end{tabular}
 \caption{Patient characteristics in the dataset}
  \label{tab:1}
\end{table*}

\subsection{Basic Machine Learning Classifiers}
We started with simple models as a baseline and progressively dived into more complex ones. This allows us to solve problems efficiently, by reasoning at the simplest useful level. In addition, this enables us to explore in which direction we should refine the more complex model to address the hard parts better. 
We have evaluated Logistic Regression (LR) and Support Vector Machine (SVM) models. These models were trained on the patient's metadata, medical history, smoking status, hospitalisation status, and symptoms.

\subsection{Convolutional Neural Network (CNN)}
With the development of deep learning, more and more deep learning methods are applied to various scenarios, such as image recognition, image classification, speech recognition, machine translation, etc. CNNs contain three main layers: convolutional layer, pooling layer, and fully connected layer. Due to the outstanding performance of neural networks, CNN has solved several complex challenges of computer vision. Therefore, a pre-trained ResNet50 convolutional neural network was used to process Mel-spectrogram images that were extracted from the audio recordings. 

The Mel-spectrogram is an effective tool to extract hidden features from audio and visualize them as an image. A CNN model can effectively extract features from images, and then complete tasks such as classification and recognition. Therefore, we use the CNN model to effectively classify the audio and to realize the accurate recognition and detection of voice.

The CNN was trained by an Adam optimizer, whose learning rate is 0.0001. The max epoch and batch size were 50 and 32, respectively. 

\subsection{Long-Short Term Memory (LSTM)}
Long Short-Term Memory (LSTM) \cite{gers2000learning} is an advanced variant of recurrent neural network(RNN) that is suited to handle sequential data and describe temporal dependencies in the data. 
\cite{sak2014long} have shown that LSTM based RNN architectures makes more effective use of model parameters to train acoustic models for large vocabulary speech recognition. Recent works have successfully employed LSTM models for disease detection (i.e. \cite{rizvi2020lstm, xue2021detection}
The intuition behind this choice is that LSTM allows the neural network to retain (and gradually forget) information about previous time instants taking advantage of the strong temporal dependency that exists between consecutive frames in the speech signals.
Our LSTM model was implemented in Python using the Keras library.

We extracted  Mel-frequency cepstral coefficients (MFCC) from the recorded audio signals. These features were used to train the LSTM model. MFCC is a fundamental feature that is widely utilized in many domains (i.e. speech recognition or emotion analysis) Its success stands in the ability to capture in a compact form and on capturing in the model the ways human hearing perceive sound \cite{logan2000mel} (i.e. humans cannot perceive frequencies over 1000 Hz). Specifically, MFCC is based on the known variations of the human ear’s critical bandwidth frequency.

Consequently, the input consists of the patient’s voice. Then, the MFCC features are extracted, and the output is passed through the LSTM model for classification.

\section{Experimental Setting}
\label{exp}
\subsection{Metadata Preprocessing}
Preprocessing is performed to account for the potential missing, incomplete or noisy data in the dataset. After data investigation, we did not have any missing values in our dataset but some columns have "none" values or "prefer not to say" entries. 
We started by categorizing the medical history of the patients in a more general categories as shown in Table \ref{tab:MED}. After that, we used one hot encoding to indicate whether a patient has a certain medical condition by using 1 if the patient has a certain medical condition and 0 otherwise. For example: a patient with high blood pressure, diabetes and cardiovascular disease. Therefore we label 1 the corresponding columns and 0 otherwise.
The same technique was used for the patient's gender. We created 3 columns for females, males, and unprovided gender. 
The smoking status includes non-smoker, prefer not to say, ex-smoker, smoked once, 1-10 cigarettes per day, 11-20 cigarettes per day, and 21+ cigarettes per day. Some smoking categories could be removed without loss of information, such as "ItOnce" can be included into "non-smoker" (as it means that the patient has tried smoking but it is not a smoker). Also, patients who smoke "21+ cigarettes per day" can be merged with the "11-20 cigarettes per day", resulting in a group of patients who smoke "11 or more cigarettes per day". One-hot encoding was also applied to the smoking status. 
As for patient's age, patients who preferred to not say their age and tested negative were removed from the dataset. Then we created age ranges instead of an exact age of a patient. An average of the age range was used, for instance, if the patient is between 40-49 years old, its age will be the average of 40-49 which is 44.5 years old. The age values were then normalized to have similar values as the values in other columns. 

The label "COVID-19 test result", encodes the patients who tested tested by 1 and patients who tested negative by 0. Similarly, for "Hospitalized" which is an indicator whether the patient is in hospital, we encode "yes" as 1 and "no" as 0.

Patient's symptoms, were also processed similarly to patient's medical history. Since most patients have no symptoms but some of them have multiple symptoms, we created a column for each symptom. If a patient has multiple symptoms, it will have the value 1 for the mentioned symptom. For example, if a patient suffers from dry cough and short breath, it will have 1 in the dry cough and short breath columns and 0 in the remaining columns of symptoms

\begin{table}[h]
    \centering
    \begin{tabular}{|P{1cm}|P{2cm}|}
    \hline
        Disease & Category\\\hline
         Hpb & high blood pressure\\\hline
         Asthma & pulmonary\\\hline
         Pnts & other\\\hline
         Longterm & other\\\hline
         Lung & pulmonary\\\hline
         Heart & cardiovascular\\\hline
         Valvular & cardiovascular\\\hline
         Cancer & Cancer\\\hline
         Diabetes & Diabetes\\\hline
         Copd & pulmonary\\\hline
         Hiv & other\\\hline
         otherHeart & cardiovascular\\\hline
         Cystic & pulmonary\\\hline
         Angina & other\\\hline
    \end{tabular}
    \caption{Medical History Categorization}
    \label{tab:MED}
\end{table}

\subsection{Feature extraction}
We extracted vocal features from the audio files that were then used into our models.  In the proposed work, the extracted features for the speech signal optimum representation are the Mel-frequency cepstral coefficients (MFCC). MFCC is a fundamental feature that is utilized in speaker and emotion recognition by virtue of the advanced representation of human auditory perception it provides. 

MEL spectrum images were also extracted. The Mel spectrum contains a short-time Fourier transform (STFT) for each frame of the spectrum (energy/amplitude spectrum), from the linear frequency scale to the logarithmic Mel-scale, the filter bank will then determine the eigenvector. Eigenvalues can be roughly expressed as the distribution of signal energy on the Mel-scale frequency. We transformed the audio recordings into Mel-spectrograms so that we can train the convolutional neural networks to distinguish between positive and negative patients. 

Finally, other additional feature were extracted such as the mean, standard deviation, maximum value, minimum value, median, kurtosis, 1st quartile, 3rd quartile, interquartile, skewness and mode of the audio signals. 

\subsection{Imbalanced dataset}
Our dataset consists of 893 voice recordings with 308 positively tested patients. First, we split our dataset into training, testing, and validation sets before balancing the data. That way, we ensure that the test dataset is as unbiased as it can be and reflects a true evaluation for our model. 
The training set has 243 negative cases and 72 positive cases. We can see that the training set is highly imbalanced, so we need to take that into consideration. Also, the testing set has 189 negative cases and 94 positive cases. Finally, the validation set has 152 negative cases and 142 positive cases. 
Given that the validation set big, we re-assigned some of positively tested patients from the validation set to the training set to increase the size of the positively test patients in training data set and thus solve the imbalance dataset problem. 
Consequently, the size of the new training set is with 214 positive and 243 negative cases, respectively. This new training set will be used as an input for all the models that are described in this paper.

\section{Experiments}
\label{expeiments}
\subsection{Experiments with neural network models}
Neural networks  have been frequently utilized to analyse data for the diagnosis of various disease. Some of the most famous classifiers in pattern recognition and classification tasks include CNN, LSTM and Generative adversarial network (GAN). To investigate the effectiveness of such networks in detecting COVID-19, we employed CNN and LSTM to identify the best model for COVID-19 classification. These models have been trained on different data inputs. The CNN was trained on MEL-spectogram images extracted from the voice recordings. Whereas, the LSTM model was fed MFCC features. 

\subsection{Experiments with basic Machine Learning models}
To test whether basic Machine Learning models improve COVID-19 prediction accuracy, two models were developed. In this paper, we examine the performance of Logistic Regression and Support Vector Machine. Thus, the proposed models were trained on the pre-processed patient's data. 

\section{Results}
\label{results}
For each experiment, several performance measures obtained during model evaluation are reported for every model. 
An overview of the performances and parameters of the used models are summarized in Table \ref{tab:overview}

\begin{table}[h]
    \centering
    \begin{tabular}{|P{0.6cm}|P{3.5cm}|P{0.8cm}|P{1cm}|P{1cm}|}
    \hline 

        Model & Parameters & Accuracy & Sensitivity & Specificity\\\hline
        LR &Input = patient's medical history, gender, smoking status, age, symptoms, hospitalized & 75\% & 70\% & 80\%\\\hline
        SVM &Input = patient's medical history, gender, smoking status, age, symptoms, hospitalized, kernel= rbf, C=1,gamma= auto & 75\% & 71\% & 79\%\\\hline
        CNN &Type = ResNet50, input = Mel-spectrogram images ,input shape=(150, 150,3), Trainable parameters $\approx$ 23M, non-trainable parameters $\approx$ 53K , loss= binary crossentropy, optimizer= adam, activation = softmax & 80\% & 79\% & 80\%,\\\hline
        LSTM & Input= MFCC features, total parameters $\approx$ 849K ,loss= binary crossentropy, optimizer= adam, activation = softmax & \textbf{89\%} & \textbf{89\%} & \textbf{83\%} \\\hline 
        
    \end{tabular}
    \caption{ A summary of the performances and parameters of the used models}
    \label{tab:overview}
\end{table}

\subsubsection{Performance Measurements}
\label{pm}
In order to better evaluate the performance of the model, we list several indicators used to evaluate the model.

\noindent \textbf{Accuracy:}
Measures the ratio between the number of correctly predicted data samples out of all the data samples.
\noindent \textbf{Sensitivity:}
Sensitivity measures the ratio of correctly predicted positive observations to the all observations in the positive class. The question it answers is: Of all the patients that truly were positive, how many did the model label as positive?

\noindent \textbf{Specificity:} is the model ability to correctly identify non-infected participants. The question answered by this metric is: Out of all the people that do not have the disease, how many received negative results?

\noindent \textbf{Confusion matrix:}
Defines a matrix summarizing the classification outcome in relation to the ground truth. Thus, the rows of each of the confusion matrices correspond to the ground-truth labels, and the columns illustrate the predicted labels.

\noindent \textbf{ROC (receiver operating characteristic) curve:}
is a graph showing the performance of a classification model at all classification thresholds. The plot shows the diagnostic ability of the binary classification models as a trade-off between sensitivity and specificity. Models with ROC curves closer to the top-left corner indicate a better performance.
 
\subsection{Experimental Results of the deep learning models}
This section highlights the importance of deep learning models in the implementation of sophisticated diagnostic systems. Two proposed models were assessed using two types of neural networks: CNN, and LSTM. The performance of each model is illustrated in the next sections. 

\subsubsection{CNN model}
Experiments were carried out to investigate the performance of a CNN model on Mel-spectogram images. As demonstrated in Table \ref{tab:CNNcm}, this model performs well since 50 of the 63 positively tested patients (about 79 percent) are identified accurately as positive. Similarly, 60 of 75 patients (80\% of the patients) who tested negative are expected to be COVID-19 free.

\begin{table}[h]
    \centering
    \begin{tabular}{|P{1cm}|P{1cm}|P{1cm}|}
    \hline
                & Negative & Positive\\\hline
        Negative& 60  & 15   \\\hline
        Positive & 13  & 50  \\\hline
    \end{tabular}
    \caption{Confusion Matrix for CNN Model}
    \label{tab:CNNcm}
\end{table}



\begin{figure}[h!]
   \centering
  \includegraphics[width=0.8\columnwidth]{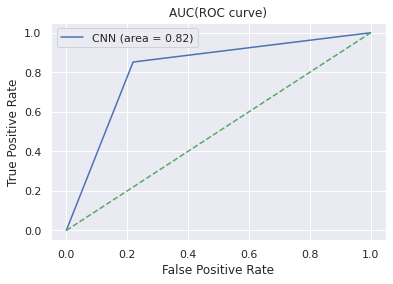}
  \caption{ROC curve for CNN Model}
  \label{fig:cnnLR}
\end{figure}

Figure \ref{fig:cnnLR} shows the ROC curve of the CNN classifier. The curve shows good performance suggesting that the CNN  model is quite accurate in detecting features from MEL-spectorgram images, that assisted in the diagnosis of patient voice recordings.

Finally, the sensitivity and specificity of our model was 79\% and 80\%, respectively.  Sensitivity of 79\% is a measure of how well the model could correctly identify patients with COVID-19 disease. Also, the high specificity indicates how well the model can correctly identify people without the disease.

\subsubsection{LSTM model}
To test the performance of the LSTM model, we ran an experiment with the MFCC features extracted from the audio recordings. According to Table \ref{tab:LSTMcm}, it is apparent that our model correctly diagnosed about 88\%  percent of our positive cases. Similarly, around 89\% of non-infected individuals were predicted to be negative.

\begin{table}[h]
    \centering
    \begin{tabular}{|P{1cm}|P{1cm}|P{1cm}|}
    \hline
                & Negative & Positive\\\hline
        Negative& 67  & 8  \\\hline
        Positive & 7  & 56  \\\hline
    \end{tabular}
    \caption{Confusion Matrix for LSTM Model}
    \label{tab:LSTMcm}
\end{table}



\begin{figure}[h!]
   \centering
  \includegraphics[width=0.8\columnwidth]{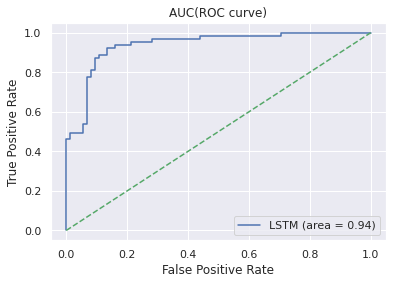}
  \caption{ROC curve for LSTM Model}
  \label{fig:LSTMLR}
\end{figure}

Figure \ref{fig:LSTMLR} shows the ROC curve for the LSTM classifier. The curve shows good performance. The ROC curve also reveals whether or not our model can discriminate between classes. Consequently, the high AUC we obtained suggests that LSTM is very good at predicting zero classes as zero and one classes as one. In other words, the higher the AUC, the better the model distinguishes between patients with and without the illness.

Lastly, we obtained with the LSTM, sensitivity and specificity of 89\%. Such high sensitivity means that there are few false negative results, and thus fewer cases of disease are missed. In addition, the high specificity of the LSTM means that the model is able to accurately designate an individual who does not have a disease as negative.

\subsection{Experimental Results of basic Machine Learning models}
We trained two basic classifiers to predict COVID-19 test result based on demographics, smoking status, symptoms, and some clinical information. 

\subsubsection{Logistic regression}
According to Table \ref{tab:lrcm}, the LR model correctly recognised 79\% of the positive cases. In addition, the model accurately classified 70\% of our negative samples.

\begin{table}[h]
    \centering
    \begin{tabular}{|P{1cm}|P{1cm}|P{1cm}|}
    \hline
        & Negative & Positive\\\hline
        Negative& 60 & 15 \\\hline
        Positive & 19 & 44\\\hline
    \end{tabular}
    \caption{Confusion Matrix for Logistic Regression Model}
    \label{tab:lrcm}
\end{table}



\begin{figure}[h!]
   \centering
  \includegraphics[width=0.8\columnwidth]{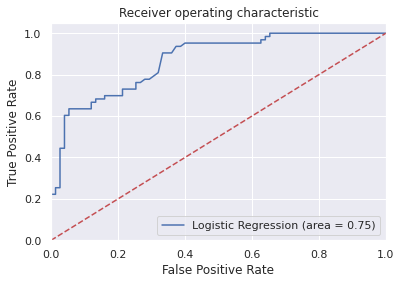}
  \caption{ROC curve for Logistic Regression Model}
  \label{fig:rocLR}
\end{figure}

As depicted by Figure \ref{fig:rocLR}, the LR model produced
a curve that is relatively close to the top-left corner, indicating an acceptable performance. 

Finally, the sensitivity and specificity of our model was
70\% and 80\%, respectively. 

\subsubsection{SVM model}

Table \ref{tab:SVMcm} shows 45 of 63 individuals who were classified as having a positive test result, they actually have the disease. Furthermore, 59 of the 75 healthy individuals were correctly diagnosed as being COVID-19-free.

\begin{table}[h]
    \centering
    \begin{tabular}{|P{1cm}|P{1cm}|P{1cm}|}
    \hline
        & Negative & Positive\\\hline
        Negative& 59 & 16 \\\hline
        Positive & 18  & 45\\\hline
    \end{tabular}
    \caption{Confusion Matrix for SVM Model}
    \label{tab:SVMcm}
\end{table}



\begin{figure}[h!]
   \centering
  \includegraphics[width=0.8\columnwidth]{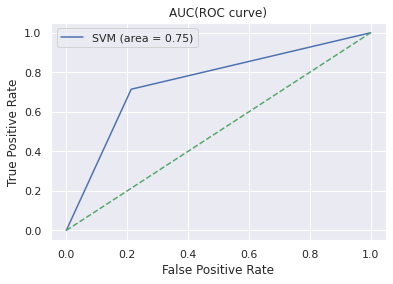}
  \caption{ROC curve for SVM Model}
  \label{fig:SVMLR}
\end{figure}

The ROC curve obtained with SVM is similar to the LR model's curve. The resulting curve, as illustrated in Figure \ref{fig:SVMLR}, demonstrates satisfactory performance.

SVM obtained a sensitivity of 70\% and a specificity of 80\%. 

We generated moderately to significantly accurate predictive models for COVID-19 detection based on voice features, demographics, symptoms, medical history, and other variables. 
The LSTM classifier is the algorithm with the highest performance. CNN turned out to be the second best. Given the accuracy of CNN, the model appears to distinguish voice features from the images of the MEL-spectogram images provided in input. LR and SVM classifiers also have a high accuracy. As demonstrated in Table \ref{tab:overview}, the detection sensitivity of the LSTM is higher than the other models, achieving 89\%. As shown in Table \ref{tab:overview}, detection specificity rose with LSTM, reaching 89\%. 

\section{Conclusion}
\label{con}
Non-intrusive and easy to collect voice parameters may provide valuable information COVID-19 diagnosis. Using voice parameters extracted from audio recordings we defined a predictive model that can classify COVID-19 patients as positive and negative. This exploratory work suggest hat the best performance is obtained with an LSTM model trained on voice features extracted as mel-frequency-cepstral-coefficients (MFCCs).
In future works we plan to further explore which features extracted from the voice signals are the most useful to distinguish COVID-19 patients. Furthermore we plan to analyse the performance of the models with a larger dataset. We will seek to externally validate the models with other voice COVID-19 datasets collected in other circumstances. More research needed to evaluate if the created models are able to distinguish between COVID-19 and another respiratory conditions.

\ifCLASSOPTIONcompsoc

\bibliographystyle{IEEEtran}
\bibliography{covid-19Detection}

\begin{thebibliography}{10}
\providecommand{\url}[1]{#1}
\csname url@samestyle\endcsname
\providecommand{\newblock}{\relax}
\providecommand{\bibinfo}[2]{#2}
\providecommand{\BIBentrySTDinterwordspacing}{\spaceskip=0pt\relax}
\providecommand{\BIBentryALTinterwordstretchfactor}{4}
\providecommand{\BIBentryALTinterwordspacing}{\spaceskip=\fontdimen2\font plus
\BIBentryALTinterwordstretchfactor\fontdimen3\font minus
  \fontdimen4\font\relax}
\providecommand{\BIBforeignlanguage}[2]{{%
\expandafter\ifx\csname l@#1\endcsname\relax
\typeout{** WARNING: IEEEtran.bst: No hyphenation pattern has been}%
\typeout{** loaded for the language `#1'. Using the pattern for}%
\typeout{** the default language instead.}%
\else
\language=\csname l@#1\endcsname
\fi
#2}}
\providecommand{\BIBdecl}{\relax}
\BIBdecl

\bibitem{worldometer2022covid}
Worldometer, ``Covid-19 coronavirus outbreak,'' 2022.

\bibitem{lai2020severe}
C.-C. Lai, T.-P. Shih, W.-C. Ko, H.-J. Tang, and P.-R. Hsueh, ``Severe acute
  respiratory syndrome coronavirus 2 (sars-cov-2) and coronavirus disease-2019
  (covid-19): The epidemic and the challenges,'' \emph{International journal of
  antimicrobial agents}, vol.~55, no.~3, p. 105924, 2020.

\bibitem{for2021science}
N.~C. for Immunization \emph{et~al.}, ``Science brief: Sars-cov-2 and surface
  (fomite) transmission for indoor community environments,'' in \emph{CDC
  COVID-19 Science Briefs [Internet]}.\hskip 1em plus 0.5em minus 0.4em\relax
  Centers for Disease Control and Prevention (US), 2021.

\bibitem{ningthoujam2020covid}
R.~Ningthoujam, ``Covid 19 can spread through breathing, talking, study
  estimates,'' \emph{Current medicine research and practice}, vol.~10, no.~3,
  p. 132, 2020.

\bibitem{han2022sounds}
J.~Han, T.~Xia, D.~Spathis, E.~Bondareva, C.~Brown, J.~Chauhan, T.~Dang,
  A.~Grammenos, A.~Hasthanasombat, A.~Floto \emph{et~al.}, ``Sounds of
  covid-19: exploring realistic performance of audio-based digital testing,''
  \emph{NPJ digital medicine}, vol.~5, no.~1, pp. 1--9, 2022.

\bibitem{stasak2021automatic}
B.~Stasak, Z.~Huang, S.~Razavi, D.~Joachim, and J.~Epps, ``Automatic detection
  of covid-19 based on short-duration acoustic smartphone speech analysis,''
  \emph{Journal of Healthcare Informatics Research}, vol.~5, no.~2, pp.
  201--217, 2021.

\bibitem{de2021covid}
S.~de~la Fuente~Garcia, F.~Haider, and S.~Luz, ``Covid-19: Affect recognition
  through voice analysis during the winter lockdown in scotland,'' in
  \emph{2021 43rd Annual International Conference of the IEEE Engineering in
  Medicine \& Biology Society (EMBC)}.\hskip 1em plus 0.5em minus 0.4em\relax
  IEEE, 2021, pp. 2326--2329.

\bibitem{hassan2020covid}
A.~Hassan, I.~Shahin, and M.~B. Alsabek, ``Covid-19 detection system using
  recurrent neural networks,'' in \emph{2020 International conference on
  communications, computing, cybersecurity, and informatics (CCCI)}.\hskip 1em
  plus 0.5em minus 0.4em\relax IEEE, 2020, pp. 1--5.

\bibitem{mehrabadi2021detection}
M.~A. Mehrabadi, S.~A.~H. Aqajari, I.~Azimi, C.~A. Downs, N.~Dutt, and A.~M.
  Rahmani, ``Detection of covid-19 using heart rate and blood pressure: Lessons
  learned from patients with ards,'' in \emph{2021 43rd Annual International
  Conference of the IEEE Engineering in Medicine \& Biology Society
  (EMBC)}.\hskip 1em plus 0.5em minus 0.4em\relax IEEE, 2021, pp. 2140--2143.

\bibitem{liang2017vibration}
J.-S. Liang and K.~Wang, ``Vibration feature extraction using audio spectrum
  analyzer based machine learning,'' in \emph{2017 International conference on
  information, Communication and Engineering (ICICE)}.\hskip 1em plus 0.5em
  minus 0.4em\relax IEEE, 2017, pp. 381--384.

\bibitem{brown2020exploring}
C.~Brown, J.~Chauhan, A.~Grammenos, J.~Han, A.~Hasthanasombat, D.~Spathis,
  T.~Xia, P.~Cicuta, and C.~Mascolo, ``Exploring automatic diagnosis of
  covid-19 from crowdsourced respiratory sound data,'' \emph{arXiv preprint
  arXiv:2006.05919}, 2020.

\bibitem{vahedian2021you}
A.~Vahedian-Azimi, A.~Keramatfar, M.~Asiaee, S.~S. Atashi, and M.~Nourbakhsh,
  ``Do you have covid-19? an artificial intelligence-based screening tool for
  covid-19 using acoustic parameters,'' \emph{The Journal of the Acoustical
  Society of America}, vol. 150, no.~3, pp. 1945--1953, 2021.

\bibitem{despotovic2021detection}
V.~Despotovic, M.~Ismael, M.~Cornil, R.~Mc~Call, and G.~Fagherazzi, ``Detection
  of covid-19 from voice, cough and breathing patterns: Dataset and preliminary
  results,'' \emph{Computers in Biology and Medicine}, vol. 138, p. 104944,
  2021.

\bibitem{nassif2022covid}
A.~B. Nassif, I.~Shahin, M.~Bader, A.~Hassan, and N.~Werghi, ``Covid-19
  detection systems using deep-learning algorithms based on speech and image
  data,'' \emph{Mathematics}, vol.~10, no.~4, p. 564, 2022.

\bibitem{chang2021covnet}
Y.~Chang, X.~Jing, Z.~Ren, and B.~W. Schuller, ``Covnet: A transfer learning
  framework for automatic covid-19 detection from crowd-sourced cough sounds,''
  \emph{Frontiers in Digital Health}, vol.~3, 2021.

\bibitem{aly2022pay}
M.~Aly, K.~H. Rahouma, and S.~M. Ramzy, ``Pay attention to the speech: Covid-19
  diagnosis using machine learning and crowdsourced respiratory and speech
  recordings,'' \emph{Alexandria Engineering Journal}, vol.~61, no.~5, pp.
  3487--3500, 2022.

\bibitem{fagherazzi2021voice}
G.~Fagherazzi, A.~Fischer, M.~Ismael, and V.~Despotovic, ``Voice for health:
  The use of vocal biomarkers from research to clinical practice,''
  \emph{Digital biomarkers}, vol.~5, no.~1, pp. 78--88, 2021.

\bibitem{solana2021analysis}
G.~Solana-Lavalle and R.~Rosas-Romero, ``Analysis of voice as an assisting tool
  for detection of parkinson's disease and its subsequent clinical
  interpretation,'' \emph{Biomedical Signal Processing and Control}, vol.~66,
  p. 102415, 2021.

\bibitem{almor1999alzheimer}
A.~Almor, D.~Kempler, M.~C. MacDonald, E.~S. Andersen, and L.~K. Tyler, ``Why
  do alzheimer patients have difficulty with pronouns? working memory,
  semantics, and reference in comprehension and production in alzheimer's
  disease,'' \emph{Brain and language}, vol.~67, no.~3, pp. 202--227, 1999.

\bibitem{wroge2018parkinson}
T.~J. Wroge, Y.~{\"O}zkanca, C.~Demiroglu, D.~Si, D.~C. Atkins, and R.~H.
  Ghomi, ``Parkinson’s disease diagnosis using machine learning and voice,''
  in \emph{2018 IEEE Signal Processing in Medicine and Biology Symposium
  (SPMB)}.\hskip 1em plus 0.5em minus 0.4em\relax IEEE, 2018, pp. 1--7.

\bibitem{schuller2021interspeech}
B.~W. Schuller, A.~Batliner, C.~Bergler, C.~Mascolo, J.~Han, I.~Lefter,
  H.~Kaya, S.~Amiriparian, A.~Baird, L.~Stappen \emph{et~al.}, ``The
  interspeech 2021 computational paralinguistics challenge: Covid-19 cough,
  covid-19 speech, escalation \& primates,'' \emph{arXiv preprint
  arXiv:2102.13468}, 2021.

\bibitem{hochreiter1997long}
S.~Hochreiter and J.~Schmidhuber, ``Long short-term memory,'' \emph{Neural
  computation}, vol.~9, no.~8, pp. 1735--1780, 1997.

\bibitem{cortes1995support}
C.~Cortes and V.~Vapnik, ``Support-vector networks,'' \emph{Machine learning},
  vol.~20, no.~3, pp. 273--297, 1995.

\bibitem{o2015introduction}
K.~O'Shea and R.~Nash, ``An introduction to convolutional neural networks,''
  \emph{arXiv preprint arXiv:1511.08458}, 2015.

\bibitem{mcculloch1943logical}
W.~S. McCulloch and W.~Pitts, ``A logical calculus of the ideas immanent in
  nervous activity,'' \emph{The bulletin of mathematical biophysics}, vol.~5,
  no.~4, pp. 115--133, 1943.

\bibitem{altman1992introduction}
N.~S. Altman, ``An introduction to kernel and nearest-neighbor nonparametric
  regression,'' \emph{The American Statistician}, vol.~46, no.~3, pp. 175--185,
  1992.

\bibitem{gers2000learning}
F.~A. Gers, J.~Schmidhuber, and F.~Cummins, ``Learning to forget: Continual
  prediction with lstm,'' \emph{Neural computation}, vol.~12, no.~10, pp.
  2451--2471, 2000.

\bibitem{sak2014long}
H.~Sak, A.~Senior, and F.~Beaufays, ``Long short-term memory based recurrent
  neural network architectures for large vocabulary speech recognition,''
  \emph{arXiv preprint arXiv:1402.1128}, 2014.

\bibitem{rizvi2020lstm}
D.~R. Rizvi, I.~Nissar, S.~Masood, M.~Ahmed, and F.~Ahmad, ``An lstm based deep
  learning model for voice-based detection of parkinson’s disease,''
  \emph{Int. J. Adv. Sci. Technol}, vol.~29, no.~8, 2020.

\bibitem{xue2021detection}
C.~Xue, C.~Karjadi, I.~C. Paschalidis, R.~Au, and V.~B. Kolachalama,
  ``Detection of dementia on voice recordings using deep learning: a framingham
  heart study,'' \emph{Alzheimer's research \& therapy}, vol.~13, no.~1, pp.
  1--15, 2021.

\bibitem{logan2000mel}
B.~Logan, ``Mel frequency cepstral coefficients for music modeling,'' in
  \emph{In International Symposium on Music Information Retrieval}.\hskip 1em
  plus 0.5em minus 0.4em\relax Citeseer, 2000.

\end{thebibliography}
\end{document}